\documentclass[pra,twocolumn,showpacs]{revtex4}
\usepackage{graphicx}
\begin{document}

\bibliographystyle{unsrt}

\title{Generation of Entangled Photon Holes using Quantum Interference}
\author{T.B. Pittman}
\altaffiliation[Present address: ]{Physics Department, University of Maryland Baltimore County, Baltimore, MD 21250}
\author{J.D. Franson}
\altaffiliation[Present address: ]{Physics Department, University of Maryland Baltimore County, Baltimore, MD 21250}
\affiliation{Johns Hopkins University,
Applied Physics Laboratory, Laurel, MD 20723}

\begin{abstract}
In addition to photon pairs entangled in polarization or other variables, quantum mechanics also allows optical beams that are entangled through the absence of the photons themselves. These correlated absences, or ``entangled photon holes'', can lead to counter-intuitive nonlocal effects analogous to those of the more familiar entangled photon pairs. Here we report an experimental observation of photon holes generated using quantum interference effects to suppress the probability that two photons in a weak laser pulse will separate at an optical beam splitter.
\end{abstract}


\maketitle

    Entangled states can exhibit correlations that are stronger than those allowed by classical physics, and the possibility of their existence led Einstein and colleagues to question the completeness of quantum mechanics soon after its inception \cite{einstein35}.  Nonetheless, recent advances have allowed entanglement to be observed in a wide variety of physical systems, including atoms \cite{rowe91}, photons \cite{aspect81,shih88}, and even hybrid combinations \cite{blinov04}. Aside from their fundamental intrigue, there has been a renewed interest in the search for new types of entangled states due to their potential utility in the rapidly developing field of quantum information processing \cite{nielsenchuangbook}.

       In the optical domain, entanglement has been observed by measuring continuous variable squeezing \cite{ou92} or by measuring discrete correlations of physical properties such as polarization \cite{shih88,kwiat95}, momentum \cite{barreiro05}, or energy and emission time \cite{franson89,marcikic04} of distant single photons. Photon pairs entangled in this way have been used to explore a number of surprising phenomena including quantum teleportation \cite{bouwmeester97} and one-way quantum computing \cite{walther05}. Equally surprising, however, is a recent prediction that optical entanglement can also arise through the absence of the photon pairs themselves  \cite{franson06}. ``Entangled photon holes'' of this kind exist in an otherwise constant background of two separate optical beams.

    The concept of entangled photon holes can be understood by contrasting it with the more familiar process of parametric down-conversion (PDC) \cite{klyshkobook}. In PDC, a photon from a quasi-monochromatic laser beam pumping a nonlinear crystal can be ``split'' to produce a pair of correlated photons emerging in separate optical beams.  The sum of the photons' energies is well defined, but the time at which the pair is produced is completely uncertain.  In this scenario, the background in each individual beam is essentially empty, but there exists a uniform probability amplitude to find the two photons at any location in the emerging beams. This results in an energy-time entanglement of the photon pairs, which can be exhibited by passing the photons through distant unbalanced interferometers \cite{franson89}.

\begin{figure}[b]
\includegraphics[angle=-90,width=3.5in]{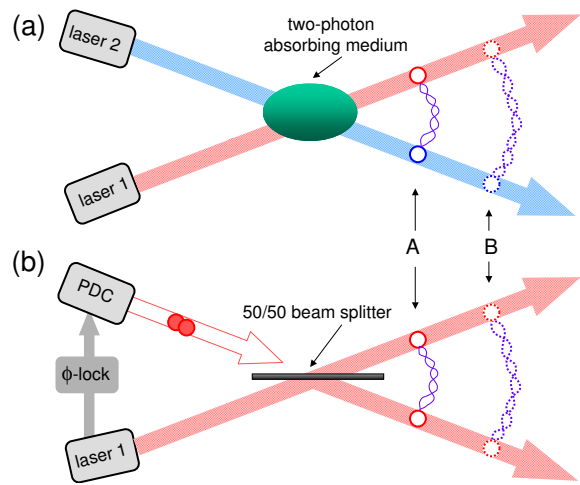}
\caption{Generation of entangled photon holes using (a) two-photon absorption, and (b) quantum interference. In (a), an idealized two-photon absorbing medium leaves an entangled photon hole pair represented by the empty circles in an otherwise constant probability amplitude background of two weak laser beams \protect\cite{franson06}.  In (b) the entangled photon holes emerge from an optical beam splitter. By phase-locking a parametric down-conversion (PDC) source to a weak laser beam, the probability to split a pair of photons from either source is suppressed by quantum interference. This eliminates the possibility of simultaneously finding a single photon in each beam, thereby creating entangled photon holes.}
\label{fig:overview}
\end{figure}

\begin{figure*}[t]
\includegraphics[angle=-90,width=5in]{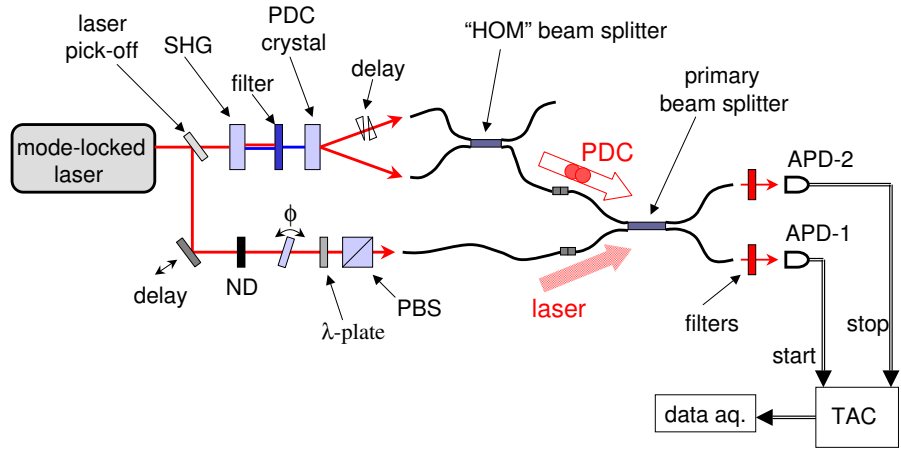}
\vspace*{-1in}
\caption{Overview of the apparatus used to observe photon holes.  An HOM interferometer \protect\cite{hong87} pumped by a frequency-doubled mode-locked Ti:Sapph laser forces pairs of PDC photons (at 780 nm) into the upper port of the primary beams splitter (a fused single-mode 3dB fiber coupler). The weak laser pulses (at 780 nm) picked off from the original laser beam are sent into the lower port of the primary beam splitter. A neutral density filter (ND) and a wave-plate followed by a polarizing beam splitter (PBS) are used to adjust the magnitude of the weak coherent state. A time-to-amplitude converter (TAC) is used to measure the difference in photon arrival times at detectors (APD-1 and 2, preceded by 3 nm bandwidth interference filters) in the beam splitter output beams. An absence of simultaneous photons reveals the correlated photon holes. The frequency doubling (SHG) and PDC crystals are 0.7 mm thick BBO. }
\label{fig:experiment}
\end{figure*}

  The idea of entangled photon holes can be loosely thought of as the ``negative image'' of PDC.  As illustrated in Figure 1(a), we consider two (different frequency) weak laser beams passing through an idealized two-photon absorbing medium that only alters the system by simultaneously removing one photon from each beam. In analogy with PDC, the two photons are known to be removed at the same time, but that time is completely uncertain.  Here the background of either emerging beam corresponds to a constant (non-zero) intensity level, but there exists a uniform probability amplitude to find a correlated absence of single-photons at any location, such as the two arbitrary points A and B shown in the figure. In analogy to the entangled photon pairs of PDC, we refer to these correlated absences in the emerging beams as entangled photon holes.

The main point of this paper is to show that entangled photon holes can be generated not only through strong two-photon absorption \cite{franson06}, but also through quantum interference effects. From a practical point of view, this moves the concept of photon holes into an experimental area that is more easily accessible with current technology. In this article, we describe experimental work in this direction.

Figure 1(b) provides a conceptual overview of the particular quantum interference method used in our experiment, in which a weak coherent state is mixed with a phase-locked PDC source on a 50/50 beam splitter. This type of setup was first used by Koashi {\em et.al.} to probe the coherence of the PDC process \cite{koashi94}. Although the results of that pioneering experiment could, in principle, be discussed along the lines of entangled photon holes, the observed quantum interference visibility was only 50\%.  In the current work, we use a substantially modified experimental implementation and observe an initial visibility of roughly 85\%. As will be described later, this could allow the entangled nature of the state to be observed, for example, through tests of Bell-type inequalities.

\begin{figure*}[t]
\vspace*{-.25in}
\includegraphics[angle=-90,width=4.5in]{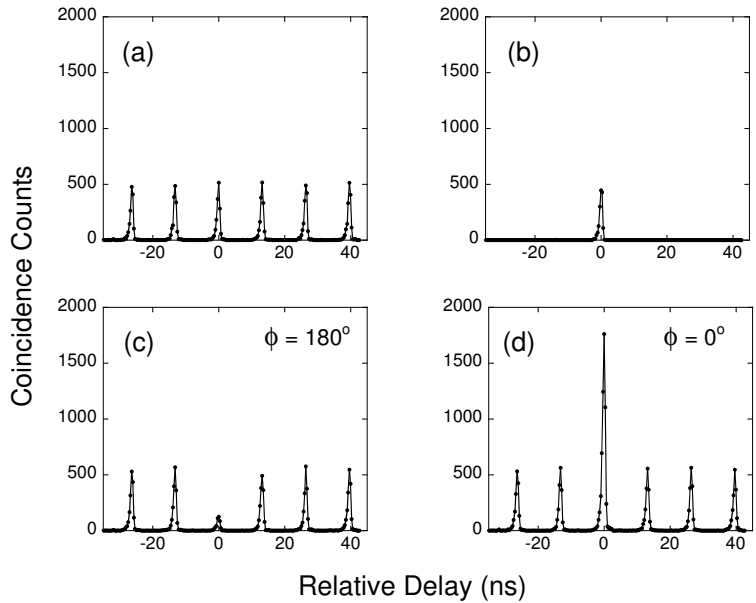}
\caption{Experimental observation of photon holes. The system was calibrated by  accumulating data with: (a) only the weak laser pulses and (b) only PDC photons. The suppression of the peak (at zero time delay) due to quantum interference in  (c) is the signature of the non-classical photon holes. The data in (d) highlights the phase-sensitive nature of the photon hole creation process. In each plot, the data accumulation time was approximately 30 seconds.}
\label{fig:data}
\end{figure*}

    In Figure 1(b) the origin of the photon holes can be understood by considering the probability of simultaneously finding one photon in each of the beam splitter output modes. This probability could arise in two ways: (I) from an amplitude corresponding to the two-photon term of the weak coherent state being split by the beam splitter, or (II) from an amplitude corresponding to the (collinear) PDC pair being split by the beam splitter \cite{koashi94}. If these amplitudes have equal magnitude, but are $180^{o}$ out of phase, they will cancel and leave a photon-hole pair. There is an equal probability amplitude for this photon hole pair to be found at any location in the output beams, just as in Figure 1(a).

    It should be noted that there are a number of closely related two-photon quantum interference effects that could be used to generate entangled photon holes in an analogous manner.  For example, the experiments of Koashi {\em et.al.} \cite{koashi93}, Resch {\em et.al.} \cite{resch01}, and Lu and Ou \cite{lu02} that used the parametric process itself, rather than a beam splitter, could be adapted for this purpose. In addition, the homodyned-PDC apparatus of Kuzmich {\em et.al.} would also suffice \cite{kuzmich00}. In analogy with the many ways entangled photon pairs can be generated, this variety of interference methods (as well as strong two-photon absorption) illustrates the flexibility of entangled photon hole state generation.

    In order to experimentally generate photon hole states in the manner of Figure 1(b), we used a fiber-based non-collinear version of the apparatus developed in  \cite{koashi94,koashi96} (see Figure 2).  The key requirement for the observation of photon holes in this system is that the two-photon amplitudes (eg. (I) and (II) ) impinging on the primary beam splitter are indistinguishable. For continuous-wave beams, this would require that the detectors used to observe the photon holes have a resolution time shorter than the coherence time of the background photons \cite{pfleegor67}. Because this is difficult to achieve \cite{lu02}, we work with ultrashort ($\sim$150 fs) pulsed light beams and slower detectors to achieve analogous results. This technique has been successfully used in most recent experiments involving multi-photon interference from independent light sources  (see, for example \cite{bouwmeester97}). The consequence in our experiment is that the background in which the photon holes appear is pulsed, rather than continuous (as was illustrated in Figure 1(b)), but the interpretation of the effects is otherwise the same.

    As shown in Figure 2, the PDC source and the weak coherent state ultimately originate from the same mode-locked pulsed laser, which facilitated control of the relative-phase of the two sources.  The PDC photons were coupled into a fiber beam splitter to implement the well known Hong-Ou-Mandel bunching effect \cite{hong87}, which forces both photons to exit via the same output port. These PDC photon pairs were then fed into the upper port of the primary beam splitter \cite{homvis}, while the weak coherent state was fed into the lower port in accordance with Figure 1(b). In this arrangement both of the two-photon amplitudes of interest corresponded to photons with identical wavelengths (780 nm) and polarizations, and spatial mode-matching was ensured by the use of single-mode fiber components. The magnitude of the weak coherent state was adjusted so that its two-photon component matched the PDC pair production rate, while the relative phase $\phi$ between the two sources could be controlled by a tilting optic in one of the beams.

    The emergence of photon holes was observed by detecting the remaining background photons with a single-photon detector in each outgoing beam. In the ideal case, each individual detector would register a constant background rate, while the correlated photon holes would be manifest through an absence of any simultaneous detections.  This could be seen by sending the outputs of the detectors to a standard ``start-stop'' data acquisition system that was used to record a histogram of the number of joint-detections (coincidence counts) as a function of the difference between detection times.

    A sample of the data collected in this manner is shown in Figure 3.  For Figure 3(a), the PDC source was blocked, so that the input to the primary beam splitter was simply that of the weak coherent state. As would be expected from a Poissonian distribution, the probability of finding a single photon in each beam at the same time is the same as two different times, so the data shows a series of constant-height peaks spaced by the repetition rate (76 MHz) of the mode-locked laser. For the data of Figure 3(b), the weak coherent state was blocked so that the input was only from the PDC source.  The single peak at zero time delay is due to the fact that the down-converted photons are produced at the same time.  Note that the magnitudes of the peaks in 3(a) and 3(b) are roughly the same.

    The main result of this paper is shown in Figure 3(c). For this data, both inputs to the primary beam splitter were open, and the relative phase $\phi$ was adjusted to equal $180^{o}$. In stark contrast to a simple incoherent addition of the peaks of 3(a) and 3(b), we see a suppression of the peak of interest (at zero time delay) due to destructive quantum interference. This corresponds to a significantly reduced probability of finding a single photon in each beam at the same time which, in this arrangement, is the signature of the photon holes.

    To further illustrate the phase-sensitive nature of the photon hole pair generation, Figure 3(d) shows the data accumulated when the relative phase $\phi$ was adjusted to be $0^{o}$. This represents the opposite of photon hole generation because the two-photon amplitudes of interest constructively interfere and increase the probability of simultaneously finding one photon in each beam.  A comparison of the zero time-delay peaks in Figures 3(c) and 3(d) indicates a quantum interference visibility of roughly 85\% \cite{drifts}.

    It is important to emphasize that, although the histogram of Figure 3(c) is similar in appearance to that which would typically be obtained by splitting a conventional antibunched beam, the statistics of that type of state are substantially different than the photon hole states of interest here.  For example, the photon number distribution of either single beam comprising a photon hole state resembles that of a coherent state, while any further splitting of an antibunched beam produces antibunched states. In addition, we note that entangled photon hole states are quite different than states produced by ``hole burning in Fock space" \cite{baseia98}, and are not the same as two-mode single-photon states of the form $|0,1\rangle+|1,0\rangle$.

    Although the data shown in Figure 3 provides evidence of photon hole generation, it should be emphasized that it does not explicitly demonstrate the entangled nature of the state. As described in reference \cite{franson06}, one way to do this would be to test Bell's inequality by sending the output beams to two distant unbalanced interferometers \cite{franson89}. In this case, the existence of the photon holes would prevent contributions to the coincidence counting rate from amplitudes corresponding to both background photons travelling the longer or shorter paths, while the required coherence between the remaining two-photon amplitudes originates from the properties of the mode-locked laser, which is known to have pulse-to-pulse phase stability over very long durations \cite{salehteichbook}. The initial visibilty of 85\% observed here should allow a violation of Bell's inequality in such an arrangement.

  The unique properties of entangled photon holes may open new doors in quantum communication and computation, as well as fundamental quantum studies. In contrast to entangled photon pairs, the holes exist in an optical background that may be less susceptible to decoherence, while still allowing nonclassical effects. In this article, we have shown that photon hole states can be generated through quantum interference effects, as well as two-photon absorption. Although the data reported here provided a first look at this new type of entangled state, open issues that remain include a more robust experimental generation of photon holes through two-photon absorption, and the theoretical development of convenient creation operators to describe them.

This work was supported by ARO, DTO and IR\&D funding.




\begin{thebibliography}{50}
\bibitem{homvis} With this particular source, we are able to observe HOM visibilities as high as 99\%, which implies that the PDC photons are nearly always sent to primary beam splitter in pairs.
\bibitem{drifts} Because active phase locking was not applied, thermal drifts of $\phi$ over the data collection period reduced the visibility from a maximum of 90\% to the value of 85\% inferred from Figure 3.
\end{thebibliography}
\end{document}